\documentclass[twocolumn,notitlepage,prl,superscriptaddress,longtable]{revtex4-1}
\usepackage{mhchem}
\usepackage{amsthm}
\usepackage{amsmath}
\usepackage{amssymb}
\usepackage{mathdots}
\usepackage{graphicx}
\usepackage{mathrsfs}
\usepackage{longtable}
\makeatletter
\@ifundefined{textcolor}{}
{%
 \definecolor{BLACK}{gray}{0}
 \definecolor{WHITE}{gray}{1}
 \definecolor{RED}{rgb}{1,0,0}
 \definecolor{GREEN}{rgb}{0,1,0}
 \definecolor{BLUE}{rgb}{0,0,1}
 \definecolor{CYAN}{cmyk}{1,0,0,0}
 \definecolor{MAGENTA}{cmyk}{0,1,0,0}
 \definecolor{YELLOW}{cmyk}{0,0,1,0}
}

\usepackage{braket}
\usepackage[backref=true,
bookmarksnumbered=true,
bookmarks=true,
bookmarksopen=true,
colorlinks=true,
citecolor=blue,
linkcolor=blue,
anchorcolor=green,
urlcolor=blue,unicode=false]{hyperref}
\renewcommand{\v}[1]{\ensuremath{\mathbf{#1}}} % for vectors
 
\let\baraccent=\= % rename builtin command \= to \baraccent
\renewcommand{\=}[1]{\stackrel{#1}{=}} % for putting numbers above =

\makeatother
\begin{document}
\title{Floquet second-order topological insulators from nonsymmorphic space-time symmetries}
\author{Yang Peng}
\email{yangpeng@caltech.edu}
\affiliation{Institute of Quantum Information and Matter and Department of Physics,California Institute of Technology, Pasadena, CA 91125, USA}
\affiliation{Walter Burke Institute for Theoretical Physics, California Institute of Technology, Pasadena, CA 91125, USA}
\author{Gil Refael}
\affiliation{Institute of Quantum Information and Matter and Department of Physics,California Institute of Technology, Pasadena, CA 91125, USA}

\begin{abstract}
We propose a systematic way of constructing Floquet second-order topological insulators (SOTIs)
based on time-glide symmetry, 
a nonsymmorphic space-time symmetry that is unique in Floquet systems.
In particular, we are able to show that the static enlarged Hamiltonian 
in the frequency domain acquires the reflection symmetry, which is inherited from the time-glide symmetry of
the original system.  As a consequence, one can construct a variety of time-glide symmetric Floquet SOTIs
using the knowledge of static SOTIs. Moreover, the time-glide symmetry only needs to be implemented
approximately in practice, enhancing the prospects of experimental realizations. 
We consider two examples,  a 2D system in class AIII and a 3D system in class A, to illustrate our ideas, and then present a general recipe for constructing Floquet SOTIs in all symmetry classes. 
\end{abstract}

\maketitle

{\em Introduction.---} 
Symmetry and topology are both at the crux of topological phases. 
Nonspatial
symmetries, i.e. the time-reversal, particle-hole and chiral symmetries,
allows a classification of topological insulators and superconductors into one of the ten Altland-Zirnbauer (AZ) symmetry classes \cite{Schnyder2008,Kitaev2009,Ryu2010,Teo2010,Chiu2016}.
When additional spatial symmetries are considered, the classification can be 
enriched, giving rise to weak topological insulators (TIs) \cite{Fu2011} protected by the lattice translational symmetry, 
as well as the topological crystalline insulators \cite{Fu2011, Ando2015},
protected by crystalline symmetries.

Recently, the ideas of utilizing crystalline symmetries were used 
to construct and understand a new family of TIs: 
the higher-order TIs \cite{Benalcazar2017, Peng2017, Langbehn2017,
Benalcazar2017s, Song2017, Schindler2018,Schindler2018b}. 
An $n$th-order TI in $d$ dimensions will have topologically protected
gapless modes that live in the $(d-n)$ dimensional boundaries, 
but all $(d-n')$ boundaries with $n'>n$ are gapped.
Thus, the conventional TIs are first order TIs
according to this definition, while the second-order TIs (SOTIs)
in two and three dimensions will host protected zero energy corner modes and gapless hinge modes 
respectively. 

On the other hand, topological phases also exist under nonequiliubrium conditions
and can be realized by time-periodic driving, known as Floquet engineering.
For instance, a Floquet TI with chiral edge modes can be brought from a static band insulator
by applying a periodic drive, such as a circularly polarized radiation or
an alternating Zeeman field \cite{Oka2009,Inoue2010,Kitagawa2011,Lindner2011,Lindner2013}.
Thus, it is natural to ask: how can  higher-order TIs
be generated with Floquet engineering? Recently, specific examples of such systems were introduced in
Ref.~\cite{Huang2018,Bomantara2018, Vega2018}.

In this work, we provide a general recipe of constructing
Floquet second-order TIs (SOTIs) in all symmetry classes, 
by making use of the dynamical nature and the time dimension in a Floquet system. 
In particular, we construct Floquet SOTIs from an 
approximate time-glide symmetry \cite{Morimoto2017},
a specific nonsymmorphic space-time crystalline symmetry \cite{Xu2018},
which is unique in a time-periodic system and has no static analog. 

The basic principle behind our construction 
is as follows. The $d-1$ dimensional boudaries in the Floquet SOTIs are essentially
stand-alone $(d-1)$ Floquet insulators from a topological
perspective, similar to their static cousin \cite{Langbehn2017}.
Hence, the topologically protected corner ($d=2$)
or hinge ($d=3$) modes naturally become domain-wall excitations
at the intersection of two gapped boundaries, if these fall into 
different topological phases. 
The approximate space-time symmetry then crucially protects such domain walls. 

The use of space-time symmetries in Floquet engineering 
of SOTIs may offer certain advantages over other recipes 
of creating SOTIs based only spatial point group symmetries.
Using external time-dependent fields may remove stringent requirements on material structure, and introduce more controllability. To wit, space-time symmetries can be induced externally,
by applying alternating fields which change directions every half a period.
Moreover, these space-time symmetries need only be approximately implemented, 
further enhancing prospects for experimental realizations.

{\em Floquet second-order topological insulators with time-glide symmetry.---} 
The corner and edge modes in a Floquet SOTI actually follow the classification
of one- and two- dimensional Floquet topological insulators, see Refs.~\cite{Roy2017,Yao2017} for example.  
Thus, we only have Floquet SOTIs in certain AZ symmetry classes,
as shown in Table~~\ref{tab:AZ-classes},
where we have listed the topological invariants for each quasienergy gap.
Note that these invariants in $d$ dimensions are exactly the same as the ones
in a Floquet topological insulator in $d-1$ dimensions.  
We will show in the following that it is possible to construct Floquet SOTI
systematically in all five nontrivial AZ classes, based on 
a single time-glide symmetry $\mathcal{M}$, which ensures the 
presence of topologically protected corner or edge states. 

\begin{table}[h]
\centering
\caption{The AZ symmetry classes are defined by the presence ($\pm 1$) or absence ($0$) of 
time-reversal $\mathcal{T}$, particle-hole $\mathcal{C}$, and chiral symmetry $\mathcal{S}$. 
The values $\pm 1$ correspond to $\mathcal{T}^2$, $\mathcal{C}^2$, or $\mathcal{S}^2$. 
The topological invariants at a particular quasienergy gap for the two-dimensional and three-dimensional Floquet SOTIs, 
which can be constructed
from time-glide symmetric Floquet topological phases, are listed
in last four columns, as well as the time-glide symmetry $\mathcal{M}$.
The symbols $\mathcal{M}_{\eta_S}$, $\mathcal{M}_{\eta_T}$, $\mathcal{M}_{\eta_C}$ and $\mathcal{M}_{\eta_T\eta_C}$
refer to a time-glide operator that squares to one and commutes ($\eta=+$) 
or anticommutes ($\eta = -$) with $\mathcal{S}$,  $\mathcal{T}$ and $\mathcal{C}$,
i.e. $\mathcal{M}\mathcal{S} = \eta_S \mathcal{S}\mathcal{M}$,
$\mathcal{M}\mathcal{T} = \eta_T \mathcal{T}\mathcal{M}$, and $\mathcal{M}\mathcal{C} =
\eta_C\mathcal{C}\mathcal{M}$.
\label{tab:AZ-classes}}
\begin{tabular}{cccccccc}
\hline
Class & $\mathcal{T}$ & $\mathcal{C}$ & $\mathcal{S}$ &\multicolumn{2}{c}{$d=2$} &\multicolumn{2}{c}{$d=3$} \tabularnewline
\hline
A & 0 & 0 & 0 & $\dots$ &$0$ & $\mathcal{M}$ & $\mathbb{Z}$\tabularnewline
AIII & 0 & 0 & 1 & $\mathcal{M}_{-}$ & $\mathbb{Z}$ & $\dots$ & 0\tabularnewline
AI & 1 & 0 & 0 & $\dots$ & 0 & $\dots$ & 0\tabularnewline
BDI & 1 & 1 & 1 & $\mathcal{M}_{+-}$ & $\mathbb{Z}$ & $\dots$ & $0$\tabularnewline
D & 0 & 1 & 0 & $\mathcal{M}_{-}$ & $\mathbb{Z}_{2}$ & $\mathcal{M}_{-}$ & $\mathbb{Z}$\tabularnewline
DIII & -1 & 1 & 1 & $\mathcal{M}_{+-}$,$\mathcal{M}_{-+}$, $\mathcal{M}_{--}$ & $\mathbb{Z}_{2}$ &
$\mathcal{M}_{+-}$,$\mathcal{M}_{--}$  & $\mathbb{Z}_{2}$\tabularnewline
AII & -1 & 0 & 0 & $\dots$ & $0$ & $\mathcal{M}_{+}$, $\mathcal{M}_{-}$ & $\mathbb{Z}_{2}$\tabularnewline
CII & -1 & -1 & 1 & $\mathcal{M}_{+-}$,$\mathcal{M}_{-+}$ & $2\mathbb{Z}$ & $\dots$ & $0$\tabularnewline
C & 0 & -1 & 0 & $\dots$ & 0 & $\mathcal{M}_{-}$,$\mathcal{M}_{+}$ & $2\mathbb{Z}$\tabularnewline
CI & 1 & -1 & 1 & $\dots$ & 0 & $\dots$ & 0\tabularnewline
\hline
\end{tabular}
\end{table}

The time-glide symmetry is a nonsymmorphic space-time symmetry, which 
comprises of a spatial reflection and a half-period time translation \cite{Morimoto2017,Xu2018}.
Without loss of generality, let us focus on the situation where the reflection plane within time glide
is perpendicular to $x$. When the symmetry acts on the Bloch Hamiltonian
$H(k_x,\v{k}_\parallel,t)$, with $k_{\parallel}$ denotes the rest Bloch momenta parallel 
to the reflection plane, we have
\begin{equation}
\mathcal{M} H(k_x,\v{k}_{\parallel},t) \mathcal{M} = H(-k_x,\v{k}_\parallel,t+T/2),
\end{equation}
where $\mathcal{M}$ is the operator implementing the time-glide symmetry, which
is both unitary and hermitian. 

A complete classification of time-glide symmetric Floquet topological insulators
and superconductors in all AZ classes has been worked out in Ref.~\cite{Morimoto2017}.
It was shown that when the edge is mapped onto itself
by the reflection part of the time glide operation,
it can host protected anomalous Floquet gapless modes \cite{Rudner2013,Nathan2015}, 
even though the classification of the Floquet system
without the time-glide symmetry is trivial. 
The existence of these anomalous Floquet modes are 
distinct from the modes protected by the spatial reflection symmetry
in topological crystaline insulators,
and are purely due to the space-time dynamical symmetry which 
has no static counterpart. 

By deploying the dynamical time-glide symmetry, we can construct intrinsically non-equilibrium Floquet SOTIs with anomalous corner or hinge modes. Our recipe follows three rules,
similar to the ones for constructing static SOTIs \cite{Langbehn2017}.
First, we require one or more pairs of system boundaries are mapped 
onto each other by the reflection part of the time-glide operation.
Second, the topological classification will be trivial when
the time-glide symmetry is broken. Third, the classification
of the corresponding AZ class in $(d-1)$ dimensions 
must be nontrivial. This guarantees the time-glide-symmetry-breaking
mass, which gaps the glide-protected boundaries, is unique. 

In Table~\ref{tab:AZ-classes}, we list all the Floquet SOTIs
that can be constructed according to the above recipe. In the rest of the manuscript, we will
construct examples of Floquet SOTI hosting anomalous Floquet
corner or hinge modes, namely the modes with quasienergies 
inside the bulk gap at the Floquet zone boundaries. 

Our construction of Floquet SOTIs uses the 
frequency-domain formulation of the Floquet problem \cite{Rudner2013}.
In this formulation, the quasienergies $\{\epsilon_j\}$ 
result from diagonalizing the enlarged Hamiltonian $\mathcal{H}(\v{k})$, whose 
matrix blocks are given by $H(\v{k},t)$ as 
$\mathcal{H}_{mm'}(\v{k}) = m\omega\delta_{mm'}\mathbb{I} + H_{m'-m}(\v{k})$
with $H_n(\v{k}) =  \frac{1}{T}\int_0^T dt\, H(\v{k},t)  e^{-in\omega t}$.
Here $\mathbb{I}$ is the identity matrix of the same dimension as $H(\v{k})$, and $m,m',n\in\mathbb{Z}$. 
Moreover, quasienergies $\epsilon_j$ and $\epsilon_j+m\omega$
describe the same physical state, and only quasienergies within a single interval of $\omega$, e.g., the ``first Floquet
zone'' with $-\omega/2<\epsilon_j<\omega/2$, are unique. 

To obtain a low-energy effective theory of 
the anomalous Floquet SOTIS, we should focus on gapless edge modes near $\epsilon=\omega/2$ (modulo $\omega$), 
similar to the static case where one assumes a Dirac-like low-energy theory.
These states would always be a result of the time-dependent drive. 
For that, we focus on $2\times 2$ block of $\mathcal{H}$ containing
the two Floquet zones shifted by $(2n+1)\omega$, with some $n\in\mathbb{Z}$, namely
\begin{equation}
\mathcal{H}_{\mathrm{eff}} = \left(\begin{array}{cc}
H_{0}+(n+\frac{1}{2})\omega & H_{2n+1}\\
H_{2n+1}^{\dagger} & H_{0} - (n+\frac{1}{2})\omega
\end{array}\right) + \frac{\omega}{2}\rho_0.
\label{eq:effective_hamiltonian}
\end{equation}
with $\rho_0$ the identity in the two Floquet-zone basis. 
This describes the situation where the bottom band of $H_0+(n+1)\omega$ crosses the top band of $H_0-n\omega$, 
and $H_{2n+1}$ opens a bulk gap at the crossing. 
The last term in Eq. (\ref{eq:effective_hamiltonian}) shifts the energy origin of the problem by $\omega/2$.
What remains of $\mathcal{H}_{\mathrm{eff}}(\v{k})$ is a reflection symmetric
system, with the effective reflection symmetry operator $\mathcal{R}_{\mathrm{eff}} = \rho_z\otimes \mathcal{M}$,
where $\rho_{x,y,z}$ are the Pauli matrices in the space of the two Floquet zones.
Hence, we have mapped a Floquet system with a time-glide symmetry to a static system with a reflection symmetry
within the effective description of the anomalous Floquet edge modes.  
%(This is actually also true when considering the full enlarged Hamiltonian $\mathcal{H}(k)$, 
%as shown in the Supplemental Material \cite{suppl}.) 

Based on the Hamiltonian  (\ref{eq:effective_hamiltonian}), we construct lattice models for harmonically driven SOTIs of the form
\begin{equation}
H(\v{k},t) = H_0(\v{k}) + H_1(\v{k}) e^{i\omega t} + H_1^\dagger(\v{k}) e^{-i\omega t},
\label{eq:classAIII_cont}
\end{equation}
which couples the upper bands of $H_0+\omega$ to the lower bands of
$H_0$, corresponding to the $n=0$ case of $\mathcal{H}_\mathrm{eff}$ of 
Eq.~(\ref{eq:effective_hamiltonian}).
$H_{0,1}(\v{k})$ respect the time-glide symmetry, as well as
the non-spatial class-appropriate AZ symmetries.

Before we proceed, it is important to note that terms that gap the anomalous 
Floquet gapless modes when time-glide symmetry is broken are odd under the time-glide operation (crucially, such terms arise when edges are not locally symmetric under the mirror element of the glide). 
Hence, the masses in the quasienergy spectra of the
two $(d-1)$ boudaries, which are connected via the time-glide operation,
will generically give rise to $(d-2)$ boundary modes. 
One can, therefore, break the time-glide symmetry and still have
protected $(d-2)$ boundary modes as long as 
the gaps of the bulk and the $(d-1)$ boundaries do not close. 
Hence, the time-glide symmetry need only be 
implemented approximately, which greatly enhances the 
prospects of an experimental realization.

{\em Two-dimensional Floquet SOTI in class AIII.---}
For a Floquet system in class AIII in 2D with a time-periodic Bloch Hamiltonian
$H(k_x,k_y,t) = H(k_x,k_y,t+T)$ of period $T$,
chiral and time-glide symmetry operators $\mathcal{S}$ and $\mathcal{M}$ obey
$\mathcal{S} H(k_x,k_y,t) =  -H(k_x,k_y,-t)\mathcal{S}$ and $\mathcal{M} H(k_x,k_y,t) = H(-k_x,k_y,t+T/2)\mathcal{M}$. Without the time-glide symmetry, 2D Floquet insulators are trivial.
When, however, the time-glide symmetry anticommutes with $\mathcal{S}$,
such systems support a $\mathbb{Z}$ classification. 

The effective Hamiltonian defined in Eq.~(\ref{eq:effective_hamiltonian})
describes a reflection symmetric system in class AIII, with 
an effective chiral symmetry $\mathcal{S}_\mathrm{eff} = \rho_x\otimes\mathcal{S}$.
When $\{\mathcal{M},\mathcal{S}\} = 0$, we
have $[\mathcal{S}_\mathrm{eff},\mathcal{R}_\mathrm{eff}]=0$, 
which leads to a $\mathbb{Z}$ topological classification 
and can give rise to helical edge modes \cite{Chiu2013,Shiozaki2014}
at the reflection symmetric edge. 

Indeed, the anomalous edge state perpendicular to the time-glide plane 
can be characterized by an edge Hamiltonian $H_{\mathrm{edge}}(k_x)=\omega/2 + 
k_x\Gamma_x$, where the edge-mode velocity was rescaled to 1,
and $\Gamma_x$ describes a number of helical modes, and  satisfies $\Gamma_x^2=1$.
Because the presence of $\mathcal{S}_\mathrm{eff}$ and $\mathcal{R}_\mathrm{eff}$, 
we have $[\Gamma_x,\mathcal{S}_\mathrm{eff}\mathcal{R}_\mathrm{eff}] = 0$.
Hence, $\Gamma_x$ and $\mathcal{S}_\mathrm{eff}\mathcal{R}_\mathrm{eff}$
can be simultaneously diagonalized. 
Suppose we can add a mass term $\Gamma_m$ that respects both the effective chiral and reflection symmetries.
Then we have $\{\Gamma_m,\mathcal{S}_\mathrm{eff}\mathcal{R}_\mathrm{eff}\} = 0$, 
indicating $\Gamma_m$ can only gap out helical modes with opposite eigenvalues of
$\mathcal{S}_\mathrm{eff}\mathcal{R}_\mathrm{eff}$.
Thus, the $\mathbb{Z}$ topological index actually counts the difference between the number
of helical edge states with positive and negative eigenvalue of 
$\mathcal{S}_\mathrm{eff}\mathcal{R}_\mathrm{eff}$. 

Let us consider, for instance, the helical states with $\mathcal{S}_\mathrm{eff}\mathcal{R}_\mathrm{eff}=1$.
The reflection operation here is effectively the same as the chiral symmetry operation, 
namely $\mathcal{R}_\mathrm{eff} = \mathcal{S}_\mathrm{eff}$. 
If we further consider a spatial configuration with an edge that breaks
the effective reflection symmetry,
a mass $\Gamma_m$ that preserves the chiral symmetry, with $\{\Gamma_m,\mathcal{R}_\mathrm{eff}\} = 0$ and $\{\Gamma_m,\Gamma_x\} = 0$,
can be added to the edge Hamiltonian.
In particular, those edges which are connected via the reflection operation
will have opposite mass. 
Since class AIII in one dimension has a $\mathbb{Z}$ topological invariant, 
the mass is unique. Thus, the intersection of two reflection-related edges corresponds
to a domain wall for the edge theory, which harbors an anomalous Floquet localized state at $\omega/2$.

A lattice model that realizes such a Floquet SOTI follows the form of
Eq.~(\ref{eq:classAIII_cont}), 
with $H_0(\v{k}) = (m-\cos k_x -\cos k_y)\tau_z + b\sigma_z$, and $H_1(\v{k}) = (\sin k_y\sigma_y - i\sin k_x)\tau_-$.
Here $\sigma_{x,y,z}$ and $\tau_{x,y,z}$ are the two sets of Pauli matrices for this 4-band model,
and $\tau_{\pm} = (\tau_x \pm i\tau_y)/2$.
The chiral and time-glide symmetries are realized by $\mathcal{S} = \tau_x\sigma_x,\,\mathcal{M}= \sigma_z$.

The corresponding effective Hamiltonian $\mathcal{H}_{\mathrm{eff}}$ of Eq.~(\ref{eq:effective_hamiltonian}) 
with $n=0$ is actually block diagonalized into two blocks with $\rho_z\tau_z = \pm 1$.
The block with $\rho_z\tau_z = 1$ is actually a trivial band insulator, whereas the one with $\rho_z\tau_z = -1$
describes a reflection symmetry topological insulator with helical modes on the edge parallel to $x$
around momentum $k_x=0$ for $(m-\omega/2)\in(0,2)$, and around momentum $k_{x}=\pi$ for $(m-\omega/2)\in(-2,0)$, 
where $b$ is numerically small. In these parameter regimes, if we cut the system such that
the two edges are mapped into each other via the reflection with respect to the time-glide plane, 
we expect to find corner modes at their intersections. 
Note that this model also has a reflection symmetry implemented by $\tau_z\sigma_z$. 
One can actually introduce an additional term $b'\tau_y$ that breaks this reflection symmetry without affecting
the corner modes, as shown in the numerics.

\begin{figure}[t]
\centering
\includegraphics[width=0.49\textwidth]{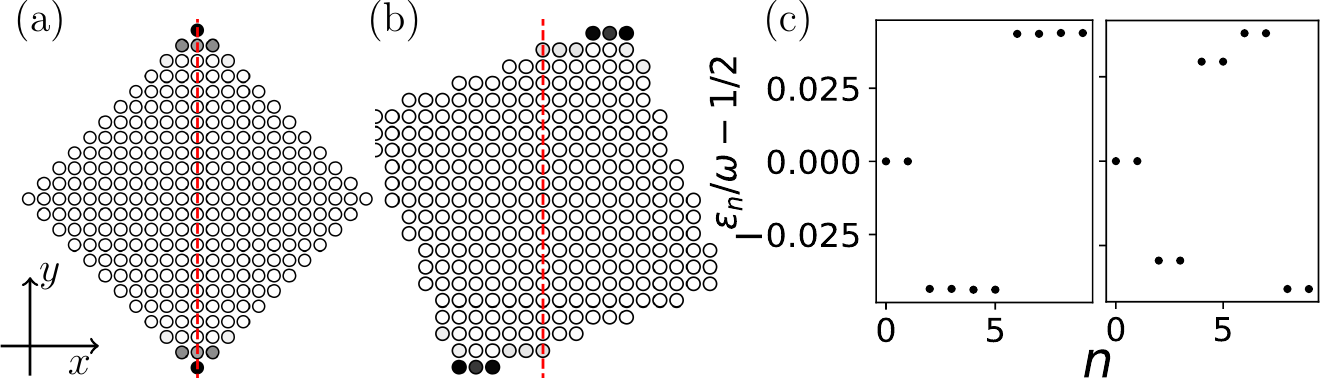}
\caption{(a,b) Support of the anomalous Floquet corner modes (darker for a larger magnitude)
at quasienergy $\omega/2$ obtained from exact diagonalization of the
enlarged Hamiltonian $\mathcal{H}$ (truncated up to $ H_0\pm 2\omega$)
for the two dimensional class AIII system with time-glide symmetry defined in
Eq.~\ref{eq:classAIII_cont}, with $\omega=6$, $m =4$, $b = 0.4$, $b'=0.8$ (reflection-symmetry-breaking term). The red dashed line indicates the 
time-glide plane. (c) 10 eigenvalues closes to $\omega/2$ for the two systems on the left. }
\label{fig:classAIII_cont}
\end{figure}

Fig.~\ref{fig:classAIII_cont}(a), depicts these states, alongside the quasienergies close to $\omega/2$ in (c). 
Even boundaries that completely break time-glide symmetry, as in (b) still gives rise to localized corner modes, which are still pinned to $\omega/2$ and separated by a smaller gap from the states at other quasienergies, see (c).
Thus, the presence of anomalous corner modes does not rely on the time-glide symmetry.

It is worth mentioning that a time-glide symmetric Floquet SOTI can also be constructed using a two-step drive, 
which may be easier to implement experimentally. For example, considering the 
system driven by two static Hamiltonians $H_{\pm}$ in the each half period,
defined as
\begin{equation}
H_{\pm} = \cos k_x \tau_x + \sin k_x \tau_y+J(\cos k_y \sigma_x\pm \sin k_y \tau_y\sigma_y), \label{3dH}
\end{equation}
where the chiral and the time-glide (with reflection of the x-direction) symmetries are realized by $\mathcal{S} = \tau_z\sigma_z$
and $\mathcal{M} = \tau_x$. 
It was shown in Ref.~\cite{Morimoto2017} that this system can also host anomalous Floquet helical edge modes protected by the time-glide
symmetry. If the system contains a pair of edges which are approximately reflect onto each other 
by the time-glide symmetry, as in
Fig.~\ref{fig:classAIII_cont}(a,b), anomalous Floquet corner modes appear at the intersections (see Supplemental Material \cite{suppl}).

{\em Three dimensional Floquet SOTI in class A.---}
Without time-glide symmetry, 3D Floquet systems in class A are 
topologically trivial. Imposing a time-glide symmetry (realized by $\mathcal{M}$) gives rise to anomalous 
Floquet surfaces modes. Consider the effective Hamiltonian $\mathcal{H}_{\mathrm{eff}}$ given in Eq.~(\ref{eq:effective_hamiltonian}).
Then $\mathcal{H}_{\mathrm{eff}}$ describes a class A system with an additional reflection symmetry
$\mathcal{R}_{\mathrm{eff}}$, which allows for a $\mathbb{Z}$ mirror Chern number
enumerating gapless surface states at reflection-symmetric surfaces \cite{Chiu2013, Shiozaki2014}.

The surface Hamiltonian describing the gapless modes on the plane normal to the z-direction can be written as
$H_{\mathrm{surface}} = \omega/2 + k_x\Gamma_x + k_y \mathcal{R}_{\mathrm{eff}}$, with $\{\Gamma_x,
\mathcal{R}_\mathrm{eff}\} = 0$. A reflection-symmetry-breaking mass $\Gamma_m$, with
$\{\Gamma_m,\mathcal{R}_\mathrm{eff}\}=0$ and $\{\Gamma_m,\Gamma_x\}=0$, will gap the surface. 
This mass is unique as class A in 2D has a nontrivial topological classification.

A 3D model also arises here
by embedding a static (x-direction) reflection-symmetric system into $\mathcal{H}_\mathrm{eff}$. 
Using the form in Eq.~(\ref{eq:classAIII_cont}),
with $H_0(\v{k}) = (m-\sum_{j=x,y,z}\cos k_j)\tau_z + b \sigma_x$
and $H_1(\v{k}) = \sum_{j=x,y,z}\sin k_j \sigma_j\tau_-$, yields a 3D Floquet SOTI, with time-glide symmetry $\mathcal{M} = \sigma_x$. 

When $(m-\omega/2) \in (1,3)$, and $b$ numerically small, the $\rho_z\tau_z = -1$
block of $\mathcal{H}_\mathrm{eff}$ (of the form of Eq.~(\ref{eq:effective_hamiltonian})) is a 3D reflection-symmetric topological crystaline insulator
in class A, with a gapless surface mode on the boundary normal to $x$. When we have two surfaces
related by the reflection symmetry, a localized hinge mode appears at the intersection of
the two surfaces. This corresponds to the anomalous Floquet modes of the full harmonically driven system. 
Similar to the class AIII case, this model also has a reflection symmetry implemented by $\tau_z\sigma_x$.
One can get rid of this symmetry by invoking $b_1\tau_x, b_2\tau_y$ etc., without
affecting the hinge modes. 

\begin{figure}[t]
\centering
\includegraphics[width=0.49\textwidth]{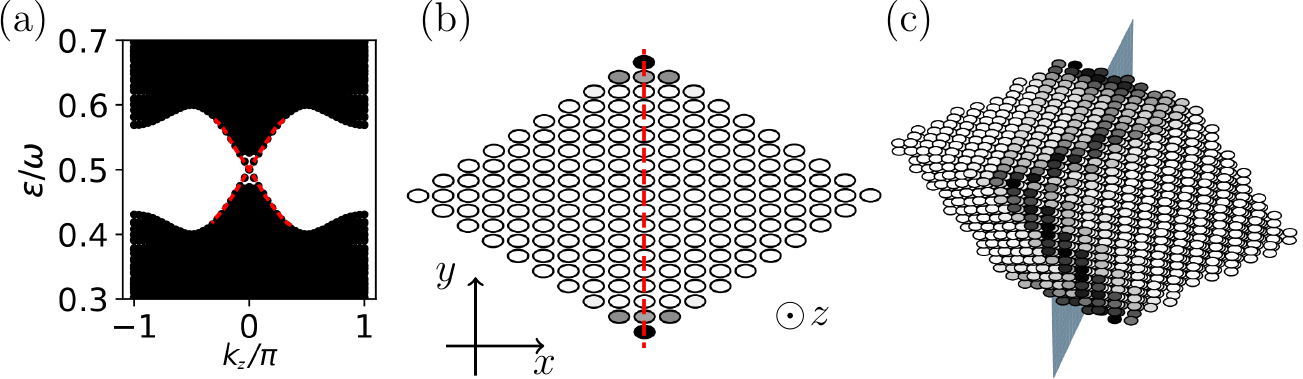}
\caption{(a) Bulk (black) and hinge (red) Floquet band structure near $\omega/2$ obtained
from exact diagonalization of the enlarged Hamiltonian $\mathcal{H}$ (truncated up to $H_0 \pm 2\omega$)
at each momentum along $z$ (periodic boundary condition), 
for the three dimensional harmonically driven Floquet system in class A, with time-glide symmetry.
The parameters are $\omega=10$, $m =7$, $b = 0.4$.
(b) Support of the anomalous Floquet hinge modes at $k_z=0$
at quasienergy $\omega/2$. 
(c) Support of the hinge modes with open boundary conditions along all directions. Here 
all surfaces breaks the reflection symmetry about the time-glide plane, which is shown in blue.}
\label{fig:classA_cont}
\end{figure}

Fig.~\ref{fig:classA_cont}(a) presents a computation of the quasienergies of the Floquet
hinge mode as a function of momentum $k_z$,
with periodic boundary conditions assumed along the $z$ direction. 
In (b), the support of the hinge mode at $k_z = 0$ was shown. 
When we consider a finite cube, where all surfaces
generically break the reflection symmetry around the time-glide plane, 
as in (c), we see the chiral Floquet hinge mode zigzags
along certain hinges of the cube, as in the  static 3D class A SOTI  \cite{Langbehn2017}.

%In fact, instead of using the harmonic drive, 
%the anomalous Floquet hinge modes can also be with discrete drives. 
%In particular, if one can take the time-glide symmetric Floquet topological insulator 
%engineered from a four-step drive introduced in Ref.~\cite{Morimoto2017},  
%and break the reflection symmetry around the time-glide plane at the boundaries, 
%the anomalous Floquet hinge would appear. We confirm this model
%explicitly in the Supplemental Material \cite{suppl}.

{\em Floquet SOTI in real symmetry classes.---}
As we claimed above, the recipe of constructing Floquet SOTIs is completely general and
can also be applied to real symmetry classes with 
time-reversal (TR) $\mathcal{T}$ and/or particle-hole (PH) $\mathcal{C}$ symmetries, 
which give rise to an effective TR
$\mathcal{T}_{\mathrm{eff}} = \rho_0\otimes\mathcal{T}$, 
or/and an effective PH $\mathcal{C}_{\mathrm{eff}} = \rho_x\otimes\mathcal{C}$
symmetries on the frequency domain effective Hamiltonian $\mathcal{H}_{\mathrm{eff}}$
defined in Eq.~(\ref{eq:effective_hamiltonian}).

For a system with time-glide symmetry $\mathcal{M}_{\eta_T, \eta_C}$ where $\eta_T,\eta_C$ 
characterizes the commutation relation between time-glide, and the TR and PH, if they exist, 
namely $\mathcal{M}\mathcal{T} = \eta_T \mathcal{T}\mathcal{M}$, $\mathcal{M}\mathcal{C} =
\eta_C\mathcal{C}\mathcal{M}$.
This determines the commutation relations between the effective reflection $\mathcal{R}_\mathrm{eff}^{\sigma_T,\sigma_C}$
and effective TR and PH, with
$\sigma_T$ and $\sigma_C$ defined similarly as
$\mathcal{R}_\mathrm{eff}\mathcal{T}_\mathrm{eff} = \sigma_T \mathcal{T}_\mathrm{eff}\mathcal{M}_\mathrm{eff}$,
$\mathcal{M}_\mathrm{eff}\mathcal{C}_\mathrm{eff} = \sigma_C\mathcal{C}_\mathrm{eff}\mathcal{M}_\mathrm{eff}$.
It is easy to show that $\sigma_T = \eta_T$ and $\sigma_C = -\eta_C$. 
In fact, Table~\ref{tab:AZ-classes} is the same as Table I in Ref.~\cite{Langbehn2017}, if we 
replace $\mathcal{R}$ by $\mathcal{M}$ while taking into account the modification of commutation relations. 
Hence, a topological property of the quasienergy gap at the Floquet zone boundary
in a time-glide symmetric Floquet system, can be obtained from analyzing
the corresponding reflection symmetric static system in the same symmetry class, according to the
mapping defined above. 

To construct time-glide symmetric Floquet SOTIs using harmonic drives, 
let us start with
a general $d$ dimensional static SOTI Hamiltonian of the form~\cite{Geier2018} $h(\v{k}) = \sum_{j=0}^d d_j(\v{k}) \Gamma_j + bB$,
where $d_0(\v{k}) = m+\sum_{j=1}^d (1-\cos k_j)$, and for $j=1,\dots d$,  $d_j(\v{k}) = \sin k_j$. Here the matrices
$\Gamma_0$ and $\Gamma_j$s are mutually anticommuting, and $B$ commutes with $\Gamma_{0,1}$ but anticommutes with the rest of the 
$\Gamma_j$s, which ensures for small $b$, that this Hamiltonian describes a topological crystaline phase with
reflection symmetry in the first coordinate. One can choose $\Gamma_0 = \tau_z$, 
and embed $h$ into the $\rho_z\tau_z = -1$ block of $\mathcal{H}_\mathrm{eff}$, 
with $m \to m - \omega/2$. This will give rise to a harmonically driven Floquet SOTI
of the form described in Eq.~(\ref{eq:classAIII_cont}) in the same AZ class of $h(\v{k})$.

{\em Conclusion.---}
In this work, we extend the second-order topological phase to the Floquet scenario. Particularly, 
we show how to systematically construct Floquet SOTIs based on time-glide symmetry, which is
a nonsymmorphic space-time symmetry unique to Floquet systems.

When a pair of boundaries in the system, which defy the mirror symmetry,
are approximately related via the reflection about the time-glide plane, a Floquet corner
or hinge modes can appear at the intersection. This can be understood in the 
frequency domain formulation of the Floquet system, by focusing on the 
effective two-by-two block of the enlarged Hamiltonian. We showed that
the this effective Hamiltonian acquires a reflection symmetry inherited 
from the time-glide symmetry, besides the AZ symmetries. 
%(Actually this is also true for the full enlarged Hamiltonian in the frequency domain, see Supplemental
%Material \cite{suppl} for details.)

Thus, the properties of the time-glide symmetric Floquet SOTI can be understood from our previous knowledge 
of the static reflection symmetry SOTI \cite{Langbehn2017}. Furthermore, we are able to systematically 
construct explicit models of harmonically driven time-glide symmetric Floquet SOTI, from Hamiltonians of static SOTIs. 
In addition to two examples (2D class AIII and a 3D class A systems), we showed that our recipe yields Floquet SOTIs in
other symmetry classes. 
%Clearly, we expect that realistic time-glide symmetric systems could be found
%using our recipe. 
Since the lattice vibrations naturally break the static reflection symmetry
while preserve the time-glide symmetry, we can expect to create Floquet SOTIs
by exciting a particular phonon mode \cite{Nova2017, Hubener2018}. On the other hand, 
the phonons can also used as a heat bath to prevent the system from heating \cite{Karthik2015,Karthik2019}. 

For other nonsymmorphic space-time symmetries that may give rise to Floquet higher-order
topological insulators, our frequency-domain analysis can be applied and
the knowledge of static systems with other crystalline symmetries can be used similarly. 
We intend to pursue these directions in our future work.  

{\em Acknowledgement.---} We acknowledge support from the Institute of Quantum In-
formation and Matter, an NSF Frontier center funded by
the  Gordon  and  Betty  Moore  Foundation,  the  Packard
Foundation.
YP is grateful to support from the Walter Burke Institute for Theoretical Physics at Caltech.
GR is grateful to the support from the  ARO  MURI  W911NF-16-1-
0361 Quantum Materials by Design with Electromagnetic
Excitation”  sponsored  by  the  U.S.  Army.

\setcounter{equation}{0}
\setcounter{figure}{0}
\newpage
\begin{widetext}
\section*{Supplemental Material}

\section*{Two dimensional Floquet SOTI in class AIII under a two-step drive}
In this section, we consider a two-step driven Floquet system first introduced in Ref.~\cite{Morimoto2017},
in which the authors showed that the model can harbor anomalous Floquet
edge modes protected by the combination of chiral symmetry and time-glide symmetry.
We show that this model describes a Floquet SOTI with anomalous Floquet corner
modes, when a pair of edges in the system are mapped onto each other
via reflection about the time-glide plane.

Within one full period in time, this system is driven by two static Hamiltonians $H_{\pm}$ in the each half
period, defined as
\begin{equation}
H_{\pm}(k_x,k_y) = \cos k_x \tau_x + \sin k_x \tau_y+J(\cos k_y \sigma_x\pm \sin k_y \tau_y\sigma_y),
\label{eq:twostep}
\end{equation}
where the chiral and the time-glide symmetries are realized by $\mathcal{S} = \tau_z\sigma_z$
and $\mathcal{M} = \tau_x$.

\begin{figure}[h]
\centering
\includegraphics[width=0.7\textwidth]{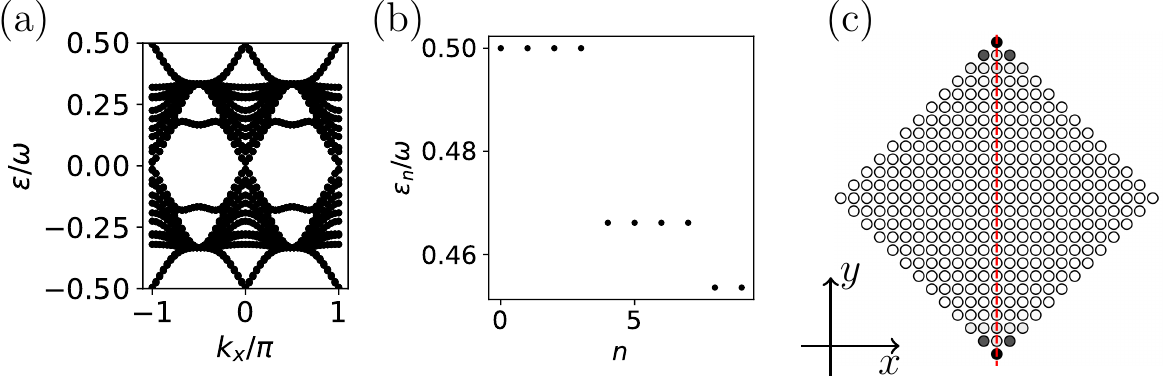}
\caption{(a) Quasienergy spectrum of the time-glide symmetric Floquet system defined in Eq.~(\ref{eq:twostep})
as a function of momentum $k_x$, when periodic boundary condition along $x$,  and open boundary condition along $y$
are imposed. There are 13 sites along $y$ in the calculation. (b) Quasienergy spectrum for the Floquet modes near $\omega/2$, when the system contains pairs of edges
which are mapped onto each other via reflection about the time-glide plane, see the geometry in (c). (c) Support
of the wave function for one of the Floquet mode at $\omega/2$. The parameters are $\omega=J=3$. }
\label{sfig:SclassAIII}
\end{figure}

In SFig.~\ref{sfig:SclassAIII}(a), we reproduce the quasienergy spectrum as the one in Ref.~\cite{Morimoto2017}, when periodic
boundary condition is imposed along $x$. We see that both at $k_x=0$  and at $k_x = \pm \pi$, the system has
gapless Floquet edges states at $\omega/2$. Hence, when we cut the system such that there are two pairs of
edges that are mapped onto each other via the reflection about the time-glide plane, anomalous
Floquet corner modes at $\omega/2$ appear, see (b). Moreover, since there are two gapless modes,
there are four degenerate corner modes at $\omega/2$, twice as many as in the ones in the harmonically
driven model introduced in the main text.  In (c), we show the support of the wave function for one of the degenerate
corner modes.

%\section*{Three dimensional Floquet SOTI in class A under a four-step drive}
%In this section, we consider a four-step driven Floquet system from Ref.~\cite{Morimoto2017} as well.
%This model has a time-glide symmetry, which gives rise to protected gapless surface modes
%when the surface is mapped on to itself under the reflection about the time-glide plane.
%The model is defined as a sum of two terms
%\begin{equation}
%H(t) = H_1(t) + H_{2}(t)
%\end{equation}
%where

\section*{Emergent reflection symmetry in the frequency-domain formulation}
In this section, we show that the full enlarged Hamiltonian in the frequency domain
aquires a reflection symmetry, whenever the original time-periodic Hamiltonian
has a time-glide symmetry. Moreover, the reflection plane coincide with the time-glide plane.

Let us write down the enlarged Hamiltonian in the frequency domain explicitly as
\begin{equation}
\mathcal{H}=\left(\begin{array}{ccccc}
\ddots\\
 & H_{0}+\omega & H_{1} & H_{2}\\
 & H_{1}^{\dagger} & H_{0} & H_{1}\\
 & H_{2}^{\dagger} & H_{1}^{\dagger} & H_{0}-\omega\\
 &  &  &  & \ddots
\end{array}\right)
\end{equation}
with
\begin{equation}
H_n(\v{k}) =  \frac{1}{T}\int_0^T dt\, H(\v{k},t)  e^{-in\omega t}.
\end{equation}

Let us first summarize how time-reversal $\mathcal{T}$, particle-hole $\mathcal{C}$ and chiral $\mathcal{S}$ symmetries
transform $H_n(\v{k})$ \cite{Yao2017}:
\begin{gather}
\mathcal{T} H_n(\v{k})\mathcal{T}^{-1} = H_n^*(-\v{k})\\
\mathcal{C} H_n(\v{k})\mathcal{C}^{-1} = -H_{-n}^*(-\v{k}) \\
\mathcal{S} H_n(\v{k})\mathcal{S}^{-1} = -H_{-n}(\v{k}).
\end{gather}

Hence, one can define the effective time-reversal $\mathscr{T}$, particle-hole $\mathscr{C}$ and chiral $\mathscr{S}$ symmetries
for the enlarged Hamiltonian $\mathcal{H}$ as
\begin{equation}
\mathscr{T}=\left(\begin{array}{ccccc}
\ddots\\
 & \mathcal{T}\\
 &  & \mathcal{T}\\
 &  &  & \mathcal{T}\\
 &  &  &  & \ddots
\end{array}\right), \quad
\mathscr{C}=\left(\begin{array}{ccccc}
 &  &  &  & \dots\\
 &  &  & \mathcal{C}\\
 &  & \mathcal{C}\\
 & \mathcal{C}\\
\dots
\end{array}\right), \quad
\mathscr{S}=\left(\begin{array}{ccccc}
 &  &  &  & \dots\\
 &  &  & \mathcal{S}\\
 &  & \mathcal{S}\\
 & \mathcal{S}\\
\dots
\end{array}\right).
\end{equation}

Let us assume the time-dependent Hamiltonian $H(\v{k},t)$ has an additional time-glide symmetry, namely
\begin{equation}
\mathcal{M}H(\v{k},t)\mathcal{M} =H(-k_x,\v{k}_{\parallel},t+T/2).
\end{equation}
When acting on $H_n(\v{k})$, the time-glide symmetry becomes
\begin{equation}
\mathcal{M}H_n(\v{k})\mathcal{M} = (-1)^{n}H_{n}(-k_x,\v{k}_\parallel).
\end{equation}
This enables us to define an effective reflection symmetry
\begin{equation}
\mathscr{R}=\left(\begin{array}{ccccc}
\ddots\\
 & \mathcal{M}\\
 &  & -\mathcal{M}\\
 &  &  & \mathcal{M}\\
 &  &  &  & \ddots
\end{array}\right),
\end{equation}
which is block diagonal with blocks alternating between $\mathcal{M}$ and $-\mathcal{M}$.
In this way, we map the original time-glide symmetric Floquet system into
a reflection symmetric static system, without changing the AZ classes.
\end{widetext}
\end{document}